\documentclass[square,comma,sort&compress]{elsart}
\usepackage{amsfonts}
\usepackage{amsmath}
\usepackage{amssymb}
\usepackage{natbib}
\usepackage{epsfig}
\usepackage{rotating}
\usepackage{graphicx}

\newcommand{\QUESTION}[1]{}
\newcommand{\ANSWER}[1]{}

\begin{document}

\runauthor{Tessarotto}
\begin{frontmatter}
\title{Phase-space Lagrangian dynamics of incompressible thermofluids}
\author[UTS,CIV,UTS1]{Marco Tessarotto}
\author[SISSA,INFN]{Claudio Cremaschini}
\author[UTS1,UTS2]{and Massimo Tessarotto}


\address[UTS]{Department of Electronics, Electrotechnics and Informatics,
University of Trieste, Italy }
\address[CIV]{Civil Protection Agency, Regione
Friuli Venezia-Giulia, Palmanova (Udine), Italy}
\address[UTS1]{Department of Mathematics
and Informatics, University of Trieste, Trieste, Italy}
\address[SISSA]{International School for Advanced Studies (SISSA), Trieste,
Italy}
\address[INFN]{National Institute of Nuclear Physics (INFN), Trieste, Italy}
\address[UTS2]{Consortium for Magnetofluid Dynamics, Trieste, Italy}

%
\thanks[maxtex]{Corresponding author: email: massimo.tessarotto@cmfd.univ.trieste.it}

\begin{abstract}
Phase-space Lagrangian dynamics in ideal fluids (i.e, continua) is
usually related to the so-called {\it ideal tracer particles}. The
latter, which can in principle be permitted to have arbitrary
initial velocities, are understood as particles of infinitesimal
size which do not produce significant perturbations of the fluid
and do not interact among themselves. An unsolved theoretical
problem is the correct definition of their dynamics in ideal
fluids. The issue is relevant in order to exhibit the connection
between fluid dynamics and the classical dynamical system,
underlying a prescribed fluid system, which uniquely generates its
time-evolution. \

The goal of this paper is to show that the tracer-particle dynamics can be {\it exactly} established
for an arbitrary incompressible fluid uniquely based on the construction of an inverse kinetic
theory (IKT) (Tessarotto \textit{et al.}, 2000-2008). As an example, the case of an
incompressible Newtonian thermofluid is here considered.


\end{abstract}

\end{frontmatter}



\section{Introduction}

A basic aspect of fluid dynamics is related to the definition of the
Lagrangian dynamics which characterizes both compressible and incompressible
fluids. The customary approach to the Lagrangian formulation is based
typically on a configuration-space description, i.e., on the introduction of
the (configuration-space)\textit{\ Lagrangian path,} $\mathbf{r}(t),$
spanning the configuration space (fluid domain) $\Omega .$ Here $\mathbf{r}%
(t)$ denotes the solution of the initial-value problem $\frac{D\mathbf{r}}{Dt%
}=\mathbf{V}(\mathbf{r},t),$ with $\mathbf{r}(t_{o})=\mathbf{r}_{o}.$ Here $%
\frac{D}{Dt}\equiv \frac{\partial }{\partial t}+\mathbf{V\cdot \nabla }$
denotes the so-called "fluid" convective derivative, $\mathbf{r}_{o}$ an
arbitrary vector belonging to $\overline{\Omega }$ and $\mathbf{V}(\mathbf{r}%
,t)$ the velocity fluid field, to be assumed continuous in $\overline{\Omega
}$ (closure of $\Omega $) and suitably smooth in $\Omega $.\ \ However, in
turbulence theory the statistical formulation for the associated joint
probability density for velocity increments requires the introduction of a
phase-space representation of suitable type \cite{Pope2000}, in which
usually the phase-space is identified with $\Gamma =\Omega \times
\mathbb{R}
^{3}$ (with closure $\overline{\Gamma }=\overline{\Omega }\times
\mathbb{R}
^{3}$), here denoted as \emph{restricted phase-space}. Therefore it is
natural to seek possible phase-space representations of this type for fluid
systems. The goal of this investigation is concerned with the formulation of
\emph{restricted-phase-space Lagrangian dynamics} in such a way that the
phase-space\emph{\ }$\Gamma $\emph{\ }coincides with the direct product
space $\Gamma =\Omega \times V,$\emph{\ }$\Omega $\emph{\ }being the fluid
domain and $V$\ (velocity space) the set $%
\mathbb{R}
^{3}$. In particular in this paper, extending the formulation
previously developed (Cremaschini \textit{et al. }
\cite{Cremaschini2008} and Marco Tessarotto \textit{et al. }
\cite{Tessarotto20083,Tessarotto20081}), we shall adopt for
this purpose a so-called \emph{phase-space inverse kinetic theory (IKT) }%
(see also Tessarotto \textit{et al.}, 2000-2008 \cite%
{Ellero2000,Ellero2004,Ellero2005,Tessarotto2006,Tessarotto2007,Tessarotto2008PP}%
). Basic feature (of such an approach) it that is relies on first
principles, i.e., classical statistical mechanics and a prescribed complete
set of fluid equation. This permits us to advance in time the relevant fluid
fields by means of \emph{phase-space Lagrangian equations} defined by the
vector field $\mathbf{X}(\mathbf{x},t),$ namely
\begin{equation}
\left\{
\begin{array}{c}
\frac{d\mathbf{x}}{dt}=\mathbf{X}(\mathbf{x},t), \\
\mathbf{x}(t_{o})=\mathbf{x}_{o},%
\end{array}%
\right.   \label{Eq.4}
\end{equation}%
where $\mathbf{x}_{o}$ is an arbitrary initial state of
$\overline{\Gamma }$ (closure of the phase-space $\Gamma )$. The
result appears relevant in particular for the following reasons:
1) the Lagrangian dynamics here determined permits to advance in
time self-consistently the fluid fields, i.e., in such a way that
they satisfy identically the required set of fluid equations. For
isothermal fluids, this conclusion is consistent with the results
indicated previously \cite{Ellero2005}; 2) the Lagrangian dynamics
takes into account the specific form of the phase-space
distribution function which advances in time the fluid fields; 3)
the theory permits an exact description of the motion of those
particles immersed in the fluid which follow the Lagrangian
dynamics (classical molecules).

Phase-space Lagrangian dynamics and particle dynamics in ideal fluids are
closely related issues. In fact both must be uniquely described via a
suitable complete set of fluid fields $\left\{ Z(\mathbf{r},t)\right\} $
which define the fluid state. This refers, in particular, to the so-called
\emph{ideal tracer particles}, for which both self-interaction produced by
the perturbations of the fluid fields generated by the same particles and
binary collisions among them are negligible (in this sense they can
therefore be intended also as "collisionless"). It is well known, however,
that in customary approaches (see for example Maxey and Riley, 1982 \cite%
{Maxey}) the equations of motion for ideal tracer particle are only known in
some approximate sense and therefore do not reproduce exactly the correct
fluid dynamics.

The purpose of this paper is to show that \textit{an exact solution }can be
reached for phase-space Lagrangian dynamics, and in particular for the
conventional tracer dynamics, based on the formulation of a suitable IKT. By
definition, an IKT must provide the complete set of fluid equations
describing the fluid, by means of velocity moments of an appropriate
phase-space probability density function (pdf). We intend to show that such
a theory can be uniquely determined in the framework of classical
statistical mechanics by invoking suitable statistical assumptions on the
IKT. In particular, we present here a theory which applies to incompressible
Newtonian fluids, including both isothermal and non-isothermal fluids \cite%
{Cremaschini2008,Tessarotto20083,Tessarotto20081}. In the following we
intend to show that customary tracer-dynamics equations due to several
authors - including Tchen (1947 \cite{Tchen1947}), Corrsin and Lumley (1956
\cite{Corrsin1956}), Buevich (1966 \cite{Buevich1966}) and Riley (1971 \cite%
{Riley 1971}) and Maxey and Riley (1982 \cite{Maxey}) -\emph{\ }are\emph{\ }%
incompatible with the exact phase-space Lagrangian formulation here obtained.

In detail the plan of the paper is as follows. First, in Section 2
previous approaches to ideal tracer particle dynamics are
summarized. Second in Sec.3 a comparison between Eulerian and
Lagrangian phase-space approaches is provided.\ Furthermore, in
Sec. 4 the IKT for incompressible thermofluids is presented. This
permits us to determine the appropriate form of the vector field $\mathbf{X}(%
\mathbf{x},t).$ Next, in Section 5 the Lagrangian formulation of IKT is
discussed in detail. The new set of phase-space Lagrangian equations are
shown to advance uniquely in time the relevant fluid fields of an
incompressible thermofluid. As a basic consequence, in Section 6 we will
derive the exact dynamics of ideal tracer particles (see below for
definition), comparing it with previous results.

\section{Previous approaches to tracer particle dynamics}

The motion of small particles (such as solid particles or droplets, commonly
found in natural\ phenomena and industrial applications) which can be
injected in a fluid with arbitrary initial velocity, in practice, may be
very different from that of the fluid. The accurate description of particle
dynamics, as they are pushed along erratic trajectories by binary collisions
(in real fluids) and by fluctuations of the fluid fields (in ideal fluids),
is fundamental to transport and mixing in turbulence \cite{Pope2000}. It is
essential, for example, in combustion processes \cite{Pope1994}, in the
industrial production of nanoparticles \cite{Pretsinis1996} as well as in
atmospheric transport, cloud formation and air-quality monitoring of the
atmosphere \cite{Villancourt2000,Weil1992}. The Lagrangian approach -
denoted as Lagrangian turbulence (LT) - has been fruitful in advancing the
understanding of the anomalous statistical properties of turbulent flows
\cite{Falkovich2001}. In particular, the dynamics of particle trajectories
has been used successfully to describe mixing and transport in turbulence
\cite{Pope1994,Shraiman2000}. Nevertheless, issues of fundamental importance
remain unresolved (see for example Refs. \cite{Sreeni2009,Beck2007} for
recent results regarding the Lagrangian view of passive scalar turbulence).\
\ In the past, the treatment of Lagrangian dynamics in turbulence was based
on stochastic models of various nature, pioneered by the meteorologist
Richardson \cite{Richardson} (see also \cite{Sreeni2009,Beck2007}). These
models, which are based on tools borrowed from the study of random dynamical
systems, typically rely - however - on experimental verification rather than
on first principles. However, in most cases there remains a lack of
experimental data to verify the reliability of such models \cite{Beck2007}.
Verification can be based, in particular, on the measurement of fluid
particle trajectories, obtained by seeding a turbulent flow with a small
number of tracer particles and following their motions with an imaging
system. \ On the other hand, the accurate evaluation of the Lagrangian
velocity in laboratory turbulence experiments requires measurements of
positions of tracer particle by using a suitable tracking system able to
resolve very short time (and spatial) scales. In practice this can be a very
challenging task since particle motions must be measured on very short time
scales.

As for the theory itself, rigorous results have been scanty, probably
because of the subject complexity. In the case of ideal or dilute real
fluids, however, particle motion is necessarily collisionless (in the sense
specified above) while the dynamics of ideal tracer particles is controlled
by the force produced on them only by the unperturbed fluid fields. Several
authors have tried in the past to derive, based on phenomenological
arguments, an approximate equation for the ideal tracer-particle dynamics,
describing the motion of a particle suspended in a non-uniform flow. Since
the original Basset-Boussinesq-Oseen (BBO) equation \cite%
{Basset,Boussinesq,Oseen}, formulated in the case of a uniform flow, several
papers have appeared proposing modifications or corrections of the same
equation for non-uniform flows (for a review see \cite{Maxey}). The first
attempt at a generalization of this type is due to Tchen, (1947 \cite%
{Tchen1947}), who considered the motion of a rigid sphere in an
incompressible isothermal Navier-Stokes (NS) fluid. Tchen \cite{Tchen1947}
derived an approximate equation of motion for a finite-size spherical
particle of radius $a$ and mass $m_{P}$ describing its dynamics in terms of
the Newtonian state of its center of mass $\left\{ \mathbf{r}(t),\mathbf{v}%
(t)\right\} $. \ His equation was later modified by Corrsin and Lumley (1956
\cite{Corrsin1956}), to take into account contributions due to pressure
gradients previously ignored, and by Buevich (1966 \cite{Buevich1966}) in
order to consider also the effect of viscous stress. The version of the
equation currently adopted by some authors (see for example Gui \textit{et
al.}, 2008 \cite{Gui2008} where it was used to investigate modifications of
turbulence) is, however, the one later developed by Maxey and Riley (1982
\cite{Maxey}) in which also the buoyancy contribution produced by the volume
displaced by the particle was taken into account. In the approximation in
which perturbations of the fluid fields produced by the particle are
negligible the equation of motion developed by Maxey and Riley reduces to:%
\begin{eqnarray}
&&m_{P}\frac{d}{dt}\mathbf{v}(t)=m_{F}\left. \frac{D\mathbf{V}(\mathbf{x},t)%
}{Dt}\right\vert _{\mathbf{x=r}(t)}-\frac{1}{2}m_{F}\left. \frac{d}{dt}%
\left\{ \mathbf{v}(t)-\mathbf{V}(\mathbf{x}(t),t)\right\} \right\vert _{%
\mathbf{x=r}(t)}-  \label{MR equation} \\
&&-\left( m_{P}-m_{F}\right) \mathbf{g}  \notag
\end{eqnarray}%
(\emph{M-R equation}), where the last contribution on the r.h.s. denotes the
so-called buoyancy effect. Here the notation is standard. Thus, $\left\{
\mathbf{V}(\mathbf{r},t),p(\mathbf{r},t)\right\} $ are respectively the
fluid velocity and pressure, $\mathbf{f}$ the volume force density and
finally $\rho _{0},\nu >0$ the constant mass density and kinematic viscosity
(with $\nu $ related to the dynamic viscosity $\mu $ by the identity $%
\mathbf{\nu =}\mu /\rho _{0}$). \ In particular, $\mathbf{f}$ can be written
\ $\mathbf{f}=-\nabla \phi +\mathbf{f}_{R},$ with $\rho _{0}\mathbf{g}%
\mathbf{=}-\nabla \phi ,$ where $\mathbf{g}$ is the constant local gravity
acceleration, $\ \phi =\rho _{0}gz$ the gravitational potential (hydrostatic
pressure) and $\mathbf{f}_{R}$ a possible non-potential force density.
Moreover, $m_{F}$ is the mass of the fluid displaced by the sphere, $\frac{d%
}{dt}\equiv \frac{\partial }{\partial t}+\mathbf{v}(t)\cdot \nabla $ the
"particle" convective derivative and $\mathbf{V}(\mathbf{r}(t),t)$ and $%
\mathbf{v}(t)-\mathbf{V}(\mathbf{r}(t),t)$ are respectively the fluid and
particle relative velocities evaluated at the position of the particle
center of mass. It should be noted, however, that also this equation is
still unsatisfactory. In fact, it is obtained by requiring that the particle
velocity remains always suitably close to the fluid one, so that
contributions due to the relative velocity - i.e., proportional to the
particle relative velocity $\mathbf{u}(t)\equiv $ $\mathbf{v}(t)-\mathbf{V}(%
\mathbf{r}(t),t)$ - are actually ignored in Eq.(\ref{MR equation}),
requiring that for all $\left\{ \mathbf{r}(t),\mathbf{v}(t)\right\} $ there
results:
\begin{equation}
\left\vert \mathbf{u}(t)\right\vert \ll \left\vert \mathbf{V}(\mathbf{r}%
(t),t)\right\vert .  \label{assumption}
\end{equation}%
The limitation appears serious because\emph{\ }the actual dynamics (i.e.,
both the velocity and the acceleration) of ideal tracer particles may be in
principle very different from that of the fluid elements. In addition, the
accurate description of ideal particle dynamics is essential both in LR and
in environmental fluid dynamics (dynamics of anthropogenic pollutants in the
atmosphere, diffusion of dusty particles, droplets, aerosol particles,
etc.). This involves, in fact, the ability to simulate tracer dynamics in a
variety of different physical conditions and in fluid flows characterized by
a turbulent behavior. Therefore,\textit{\ an open issue remains the very
definition of the dynamics of tracer particles which may be injected in an
ideal fluid with arbitrary initial velocities. }Clearly, such a formulation
- if achievable at all - should rely exclusively on first principles, i.e.,
in particular the exact validity of the fluid equations. This problem is
closely related to the formulation of phase-space approaches for fluid
systems, based on the introduction of suitable phase-space representations
of classical fluid dynamics in terms of an appropriate phase-space pdf.\ By
construction, the dynamics of a fluid is completely described by the
time-evolution of its fluid fields (which, in turn, are assumed as classical
solutions of a well-posed initial-boundary value problem defined by a
complete set of fluid equations). Therefore, unless suitable restrictions
are posed, phase-space dynamics remains intrinsically non-unique. This is
true, in particular, due to the arbitrariness in the choice of the possible
phase-space and the definition of the evolution equation for the pdf.

\section{Eulerian and Lagrangian phase-space approaches}

As it is well known, phase-space descriptions of fluids can be achieved in
principle choosing either an Eulerian or a Lagrangian point of view. \ Based
on the IKT approach for incompressible fluids earlier developed \cite%
{Ellero2000,Ellero2004,Ellero2005,Tessarotto2006,Tessarotto2007} such a
connection can be uniquely established. IKT is based on the identification
of the \emph{complete set of fluid fields} (which describe the fluid) with
velocity-moments of a suitably-defined kinetic distribution function $f(%
\mathbf{x,}t)$. The pdf is assumed to satisfy the basic principles of
classical statistical mechanics, which include in particular:

\begin{enumerate}
\item \emph{the principle of conservation of probability; }

\item a suitable \emph{correspondence principle }(Ellero \textit{et al.},
2005 \cite{Ellero2005}) \emph{\ }whereby appropriate (velocity-) moments of
the pdf can be identified with the relevant fluid fields;

\item \emph{the principle of entropy} \emph{maximization }(PEM, Jaynes, 1957
\cite{Jaynes1957});

\item and \emph{an entropic principle assuring that the statistical entropy
cannot decrease in time }(Tessarotto, 2008 \cite{Tessarotto20083}).
\end{enumerate}

The first axiom implies that the pdf must satisfy a Liouville equation,
i.e., a Vlasov-type \emph{inverse kinetic equation} (IKE, Ellero \textit{et
al.} \cite{Ellero2005}). As it is well-known, this type of kinetic equation
is, in fact, appropriate for the statistical description of particles
subject solely to mean-field interactions. This is consistent with the
assumption of an ideal fluid, i.e., a continuum in which the fluid elements
are, by definition, subject only to mean-field interactions. In such a case,
the time-evolution of the pdf is determined by the \textit{Eulerian IKE}
\begin{equation}
Lf(\mathbf{x},t)=0.  \label{Eq.1}
\end{equation}%
Here $f(\mathbf{x},t)$ denotes the Eulerian representation of the pdf, $L$
is the streaming operator $Lf\equiv \frac{\partial }{\partial t}f+\frac{%
\partial }{\partial \mathbf{x}}\cdot \left\{ \mathbf{X}(\mathbf{x}%
,t)f\right\} ,$ $\mathbf{X}(\mathbf{x},t)\equiv \left\{ \mathbf{v,F}(\mathbf{%
x},t)\right\} $ a suitably smooth vector field, while $\mathbf{v}$ and%
\textbf{\ }$\mathbf{F}(\mathbf{x},t)$ denote respectively the velocity and
an appropriate\ "mean-field" acceleration vector field. The implications of
the principle of maximum entropy \cite{Jaynes1957} and of the entropic
principle, both involving the assumption that the Boltzmann-Shannon (B-S)
entropy functional%
\begin{equation}
S(t)\equiv S(f(\mathbf{x,}t))=-\int\limits_{\Gamma }d\mathbf{x}f(\mathbf{x,}%
t)\ln f(\mathbf{x,}t)  \label{B-S entropy}
\end{equation}%
exists, have been discussed elsewhere \cite{Cremaschini2008,Tessarotto20083}
(see also Tessarotto \textit{et al.}, 2007 \cite{Tessarotto2007}). In
particular this provides a well-defined initial condition for the pdf, $f(%
\mathbf{x,}t_{o})=f_{M}(\mathbf{x,}t_{o})$ [see below Eq.(\ref{Maxwellian})]
and also a (generally non-unique) representation for the streaming operator $%
L$ (Tessarotto \textit{et al.}, 2006 \cite{Tessarotto2006}). As a main
consequence the same approach can in principle be used to determine in a
rigorous way the Lagrangian formulation for arbitrary complex fluids.
Although the choice of the phase-space $\Gamma $ is in principle arbitrary,
in the case of incompressible isothermal fluids, it is found \cite%
{Ellero2004} that it can always be reduced to the direct-product space $%
\Gamma =\Omega \times V$ (\emph{restricted phase-space})$,$ where $\Omega
,V\subseteq $ $%
\mathbb{R}
^{3},$ $\Omega $ is an open set denoted as configuration space of the fluid
(fluid domain) and $V$ is the velocity space.

\section{The case of an incompressible thermofluid}

Let us consider for definiteness an incompressible, viscous and generally
non-isentropic thermofluid (which comprises as a particular case also the
treatment of incompressible isothermal fluids earlier developed in\ Ref.\cite%
{Ellero2005}). This is described {by the fluid fields }$\left\{ Z\right\}
\equiv \left\{ \rho \geq 0,\mathbf{V},p\geq 0,T>0,S_{T}\right\} ,${\ to be
identified respectively with the mass density, the fluid velocity, pressure,
temperature and thermodynamic entropy. In the open set $\Omega $ they are
assumed to satisfy the so-called incompressible Navier-Stokes-Fourier
equations (INSFE), i.e.,}
\begin{eqnarray}
&&\rho =\rho _{o}>0,  \label{1} \\
&&\left. \nabla \cdot \mathbf{V}=0,\right.   \label{1b} \\
&&\left. \frac{D}{Dt}\mathbf{V}=\mathbf{F}_{H}-\frac{1}{\rho _{0}}\left[
\nabla p-\mathbf{f}\right] +\nu \nabla ^{2}\mathbf{V},\right.   \label{2} \\
&&\left. \frac{D}{Dt}T=\chi \nabla ^{2}T+\frac{\nu }{2c_{p}}\left( \frac{%
\partial V_{i}}{\partial x_{k}}+\frac{\partial V_{k}}{\partial x_{i}}\right)
^{2}+\frac{1}{\rho _{0}c_{p}}J\equiv K,\right.   \label{3} \\
&&\left. \frac{\partial }{\partial t}S_{T}\geq 0,\right.   \label{4}
\end{eqnarray}%
where $\rho _{o}$ is a constant,
\begin{equation}
\mathbf{F}_{H}\equiv -\frac{1}{\rho _{0}}\left[ \nabla p-\mathbf{f}\right]
+\nu \nabla ^{2}\mathbf{V}  \label{fluid acceleration}
\end{equation}%
is the fluid acceleration and $\frac{D}{Dt}=\frac{\partial }{\partial t}+%
\mathbf{V}\boldsymbol{\cdot \nabla }$ the convective derivative$.${\ These
equations are assumed to satisfy a suitable initial-boundary value problem\
(INSFE problem) so that a smooth (strong) solution exists for the fluid
fields }$\left\{ Z\right\} .${\ }Here{\ the notation is standard \cite%
{Tessarotto20083}. Thus, Eqs.(\ref{1})-(\ref{2}) together are denoted
incompressible Navier-Stokes equations (INSE) [with Eqs.(\ref{1}) and (\ref%
{1b}) representing respectively the so-called \emph{incompressibility and
isochoricity conditions }}and{\emph{\ }Eq. (\ref{2}) the forced
Navier-Stokes equation \emph{forced} \emph{Navier-Stokes equation} written
in the Boussinesq approximation], while Eq.(\ref{3}) is the \emph{Fourier
equation. }}Finally{\emph{\ }(\ref{Eq.4}) denotes the \emph{entropy
inequality}} which defines the\emph{\ second principle of thermodynamics}{.\
As a consequence, in such a case the force density }$\mathbf{f}$ {reads \ }$%
\mathbf{f}=\rho _{0}\mathbf{g}\left( 1-k_{\rho _{0}}T\right) +\mathbf{f}_{1},
$ where the first term represents the (temperature-dependent) gravitational
force density, while the second\ one ($\mathbf{f}_{1}$) the action of a
possible non-gravitational externally-produced force. Hence $\mathbf{f}$ can
be written also as $\mathbf{f}=-\nabla \phi +\mathbf{f}_{R},$where $\phi
=\rho _{0}gz$ and $\mathbf{f}_{R}=-\rho _{0}\mathbf{g}k_{\rho _{0}}T+\mathbf{%
f}_{1}$ denote\ respectively the gravitational potential (hydrostatic
pressure) and the non-potential force density. Moreover, in Eq.(\ref{3}) $K$
and $J$ are the quantities of heat generated per unit volume and unit time
by all sources and, respectively, only by the external sources. In
particular, the inequality (\ref{4}) defines the so-called \emph{2nd} \emph{%
principle} for the thermodynamic entropy $S_{T}.$ For its validity in the
following we shall assume that there results everywhere in $\overline{\Omega
}\times \overline{{I}}$%
\begin{equation}
\int_{\Omega }d\mathbf{r}\left( \chi \nabla ^{2}T+\frac{1}{\rho _{0}c_{p}}%
J\right) \geq 0,  \label{fluid with external heating}
\end{equation}%
which defines a so-called \emph{externally heated thermofluid}. In these
equations $\mathbf{g,}k_{\rho _{0}},\nu ${$,$}$\chi $ and $c_{p}$ are all
real constants which denote respectively the local acceleration of gravity,
the density thermal-dilatation coefficient, the kinematic viscosity, the
thermometric conductivity and the specific heat at constant pressure. Thus,
by taking the divergence of the N-S equation (\ref{2}), there it follows the
Poisson equation for the fluid pressure $p,$ namely $\nabla ^{2}p=-\rho
_{0}\nabla \cdot \left( \mathbf{V}\cdot \nabla \mathbf{V}\right) +\nabla
\cdot \mathbf{f},$ with $p$ to be assumed non negative and bounded in $%
\overline{\Omega }\times \overline{{I}}$. \

\subsection{Functional uniqueness of IKT}

Let us now assume that $f(\mathbf{x,}t)$ is a solution of the Eulerian
kinetic equation (\ref{Eq.1}) defined in a suitable extended phase-space $%
\Gamma \times I,$ where $I\subseteq
\mathbb{R}
$ is a suitable time interval. In such a case, we intend to show that $f(%
\mathbf{x,}t)$ [to be assume strictly positive] can be defined in such a way
that the fluid fields $\mathbf{V},p_{1}$ and $S_{T}$ can be identified with
its velocity moments $\int\limits_{%
\mathbb{R}
^{3}}d\mathbf{v}G(\mathbf{x,}t)f(\mathbf{x,}t),$ where respectively $G(%
\mathbf{x,}t)=\mathbf{v},\rho _{o}u^{2}/3,-lnf(\mathbf{x,}t),\mathbf{u\equiv
v-\mathbf{V}(\mathbf{r},}t)$ is the relative velocity and $p_{1}$ the
kinetic pressure defined as:%
\begin{equation}
p_{1}=p_{0}(t)+p-\phi +\frac{\rho _{0}}{m_{P}}T.  \label{Eq.6}
\end{equation}%
Here $p_{0}(t)$ (to be denotes as \emph{pseudo-pressure}$\mathbb{\ }$\cite%
{Ellero2005}) is an arbitrary strictly positive and suitably smooth function
defined in $I$. Moreover, $m_{P}>0$ is a constant mass, \textit{whose value
remains in principle arbitrary}. In particular it can be identified with the
average mass of the molecules forming the fluid. Finally, the thermodynamic
entropy $S_{T}$ can be identified with the Shannon statistical entropy
functional $S(f(\mathbf{x,}t)),$ provided the function $p_{0}(t)$ is a
suitably prescribed function and $f(\mathbf{x,}t)$ is strictly positive in
the whole set $\Gamma \times I.$ To reach the proof, let us first show that,
by suitable definition of the vector field $\mathbf{F}(\mathbf{x},t),$ a
particular solution of the IKE (\ref{Eq.1}) is delivered by the Maxwellian
distribution function:
\begin{equation}
f_{M}(\mathbf{x},t)=\frac{1}{\pi ^{2}v_{Th}}\exp \left\{ -\frac{u^{2}}{v_{Th}%
}\right\} ,  \label{Maxwellian}
\end{equation}%
where $v_{Th}=\sqrt{2p_{1}(\mathbf{r},t)/\rho _{0}}$ is the \emph{thermal
velocity driven by the kinetic pressure} $p_{1}(\mathbf{r},t)$. Based on the
results earlier obtained for ideal isothermal and incompressible fluids \cite%
{Ellero2005,Tessarotto2006,Tessarotto2007} and incompressible thermofluids
\cite{Tessarotto20083} the following theorem is reached:

\textbf{Theorem - IKT formulation for {INSFE}}

\emph{Let us assume that the INSFE problem admits a smooth strong solution
in }$\overline{\Gamma }\times I,$ \emph{such that the inequality (\ref{fluid
with external heating}) is fulfilled and the fluid fields }$\left\{
Z\right\} $ \emph{belong to the "minimal functional setting" (see Ref. \cite%
{Ellero2005}). Moreover, let us assume that in }$\overline{\Omega }\times I:$

\emph{1) the pdf }$f(\mathbf{x,}t)$ \emph{is suitably smooth, strictly
positive and admits the velocity moments }$G(\mathbf{x,}t)=1,\mathbf{v},\rho
_{o}\frac{u}{3}^{2},\rho _{o}\mathbf{u}\frac{u^{2}}{3},\rho _{o}\mathbf{uu;}$
\emph{thus, we denote in particular} $\mathbf{Q}=\rho _{o}\int d^{3}v\mathbf{%
u}\frac{u^{2}}{3}f$ \emph{and} $\underline{\underline{{\Pi }}}=\rho _{o}\int
d^{3}v\mathbf{uu}f\boldsymbol{;}$

\emph{2) the B-S entropy integral (\ref{B-S entropy}) exists in the time
interval }$I\subseteq
\mathbb{R}
;$

\emph{3) the inequality (\ref{fluid with external heating}) is assumed to
hold;}

\emph{4) there results identically (correspondence principle):}
\begin{equation}
\int d^{3}\mathbf{v}f(\mathbf{x,}t)=1,  \label{normalization}
\end{equation}

\begin{equation}
\int d^{3}\mathbf{vv}f(\mathbf{x,}t)=\mathbf{V(r,}t),  \label{flow
velocity}
\end{equation}

\begin{equation}
\rho _{o}\int d^{3}\mathbf{v}\frac{u}{3}^{2}f(\mathbf{x,}t)=p_{1}\mathbf{(r,}%
t),  \label{kinetic pressure}
\end{equation}

\begin{equation}
S_{T}(t)=S(f(\mathbf{x,}t)).  \label{entropy2}
\end{equation}

\emph{Then it follows that:}

\emph{T}$_{1}$\emph{) the local Maxwellian distribution }$f_{M}(\mathbf{x}%
,t) $, \emph{defined by Eq.(\ref{Maxwellian}), is a particular solution of
the inverse kinetic equation (\ref{Eq.1});}

\emph{T}$_{2}$\emph{)} \emph{\ the mean-field acceleration vector field }$%
\mathbf{F}$\emph{\ reads }%
\begin{equation}
\mathbf{F}(\mathbf{x},t;f)=\mathbf{F}_{0}+\mathbf{F}_{1}.  \label{F-1A}
\end{equation}%
\emph{The functional form of the vector fields }$\mathbf{F}_{0}\mathbf{,F}%
_{1}$\emph{\ is determined uniquely by requiring that they depend only on
the velocity moments indicated above. They read respectively:}%
\begin{equation}
\mathbf{F}_{0}\mathbf{(x,}t;f)=\frac{1}{\rho _{0}}\left[ \mathbf{\nabla
\cdot }\underline{\underline{{\Pi }}}-\mathbf{\nabla }p_{1}+\mathbf{f}_{R}%
\right] +\mathbf{D}(\mathbf{\mathbf{x,}}t)\mathbf{+}\nu \nabla ^{2}\mathbf{V,%
}  \label{F-2}
\end{equation}%
\begin{eqnarray}
\mathbf{F}_{1}\mathbf{(x,}t;f) &=&\frac{1}{2}\mathbf{u}\left\{ \frac{1}{p_{1}%
}A\mathbf{+}\frac{1}{p_{1}}\mathbf{\nabla \cdot Q}-\frac{1}{p_{1}^{2}}\left[
\mathbf{\nabla \cdot }\underline{\underline{\Pi }}\right] \mathbf{\cdot Q}%
\right\} +  \label{F-3} \\
&&+\frac{v_{th}^{2}}{2p_{1}}\mathbf{\nabla \cdot }\underline{\underline{\Pi }%
}\left\{ \frac{u^{2}}{v_{th}^{2}}-\frac{3}{2}\right\} .  \notag
\end{eqnarray}%
\emph{where}%
\begin{equation}
\mathbf{D}(\mathbf{\mathbf{x,}}t)\mathbf{=}\frac{1}{2}\left\{ \nabla \mathbf{%
V\cdot \mathbf{u+u}\cdot \nabla \mathbf{V}}\right\} ,  \label{D}
\end{equation}%
\emph{\ }%
\begin{equation}
A\equiv \frac{\partial }{\partial t}\left( p_{0}+p\right) -\mathbf{V\cdot }%
\left[ \frac{D}{Dt}\mathbf{V-}\frac{1}{\rho _{0}}\mathbf{f}_{R}\mathbf{-}\nu
\nabla ^{2}\mathbf{V}\right] +\frac{\rho _{0}K}{m_{P}}\equiv \frac{D}{Dt}%
p_{1};  \label{A}
\end{equation}

\emph{T}$_{3}$\emph{) for an arbitrary pdf }$f(\mathbf{x,}t)$ \emph{%
fulfilling assumptions 1-3 equations (\ref{normalization})-(\ref{D}) are
fulfilled identically in }$\overline{\Omega }\times I.$

\emph{PROOF }

Let us, first, prove proposition \emph{T}$_{1}$\emph{. }For this purpose,
let us assume that a strong solution of the INSFE problem exists which in
the set $\Omega \times I$ satisfies identically Eqs.(\ref{1})-(\ref{3}). In
such a case it is immediate to prove that $f_{M}(\mathbf{x},t)$ is a
particular solution of the inverse kinetic equation (\ref{Eq.1})$.$ This can
be proved either: a) by direct substitution of $f\equiv f_{M}(\mathbf{x},t)$
in Eq.(\ref{Eq.1}) (Proposition A); b) by direct evaluation of the velocity
moments of the same equation for $G(\mathbf{x},t)=1,\mathbf{v},u^{2}/3$
(Proposition B). Regarding Proposition B, we notice that the first two
moment equations coincide respectively with the isochoricity and
Navier-Stokes equations [Eqs. (\ref{1}) and (\ref{2})]. Therefore, the third
moment equation delivers the Fourier equation [Eq.(\ref{3})]. The same proof
(Proposition B) is straightforward also if $f\neq f_{M}(\mathbf{x},t).$ This
is reached again imposing the same constraint equation (\ref{normalization})
on first velocity-moment of the distribution function $f.$ \ The proof of
\emph{T}$_{2}$\emph{\ }and\emph{\ T}$_{3}$ follows in the same way by direct
evaluation of the velocity moments of the same equation for $G(\mathbf{x,}%
t)=1,\mathbf{v},\rho _{o}\frac{u}{3}^{2},\rho _{o}\mathbf{u}\frac{u^{2}}{3}%
,\rho _{o}\mathbf{uu}$. \ Q.E.D.

\section{Lagrangian formulation of IKT}

The previous results permit us to formulate in a straightforward way also
the equivalent Lagrangian form of IKE [see Eq.(\ref{1})]. The Lagrangian
formulation is achieved in two steps: a) by identifying a suitable dynamical
system, which determines uniquely the time-evolution of the kinetic
probability density prescribed by IKT. Its flow defines a family of
phase-space trajectories, here denoted as \emph{phase-space} \emph{%
Lagrangian paths} (phase-space LP's); b) by proper parametrization in terms
of these curves of the pdf and the inverse kinetic equation, the explicit
solution of the initial-value problem defined by the inverse kinetic
equation (\ref{Eq.1}) is determined. First, we notice that - in view of the
previous theorem it is obvious that the phase-space LP's must be identified
with the phase-space trajectories $\mathbf{x}(t)$ of a classical dynamical
system
\begin{equation}
\mathbf{x}_{o}\rightarrow \mathbf{x}(t)=T_{t,t_{o}}\mathbf{x}_{o}\equiv \chi
(\mathbf{x}_{o},t_{o},t)  \label{INSFE classical dynamical system}
\end{equation}%
(here denoted as \emph{INSFE dynamical system}), with $T_{t,t_{o}}$ the
corresponding evolution operator generated by the vector field $\mathbf{X}(%
\mathbf{x},t),$ to be prescribed according to the previous theorem$.$ Hence,
the initial-value problem (\ref{Eq.4}) is realized by the equations%
\begin{equation}
\left\{
\begin{array}{c}
\frac{d}{dt}\mathbf{r}(t)=\mathbf{v}(t), \\
\frac{d}{dt}\mathbf{v}(t)=\mathbf{F}(\mathbf{r}(t),t;f), \\
\mathbf{r}(t_{o})=\mathbf{r}_{o}, \\
\mathbf{v}(t_{o})=\mathbf{v}_{o},%
\end{array}%
\right.  \label{Eq.4'}
\end{equation}%
where the vector field $\mathbf{F}(\mathbf{r}(t),t;f)$ is defined by Eq.(\ref%
{F-1A}). Here, by construction $\mathbf{v}(t)$ and\ $\mathbf{F}(\mathbf{r}%
(t),t;f)$ are respectively the Lagrangian velocity and acceleration, both
spanning the vector space $%
\mathbb{R}
^{3}.$ In particular,\ $\mathbf{F}(\mathbf{r}(t),t;f),$ which is defined by
Eqs.(\ref{F-2})-(\ref{F-3}), and depends functionally on the kinetic
probability density $f(\mathbf{x},t),$ is the \emph{Lagrangian acceleration
which corresponds to an arbitrary kinetic probability density} $f(\mathbf{x}%
,t).$ \ From the theorem it follows that in the Lagrangian representation
the kinetic equation (\textit{Lagrangian IKE}) can be written in the form

\begin{equation}
J(\mathbf{x}(t),t)f(\mathbf{x}(t),t)=f(\mathbf{x}_{o},t_{o})\equiv f_{o}(%
\mathbf{x}_{o})  \label{Eq.2}
\end{equation}%
where $f_{o}(\mathbf{x}_{o})$ is{\ a suitably smooth initial pdf and }$J(%
\mathbf{x}(t),t)$ is the Jacobian $J(\mathbf{x}(t),t)=\left\vert \frac{%
\partial \mathbf{x}(t)}{\partial \mathbf{x}_{o}}\right\vert $ of the map $%
\mathbf{x}_{o}\rightarrow \mathbf{x}(t)$ which is generated by Eq.(\ref%
{Eq.4'}). It follows that the Lagrangian equation (\ref{Eq.2})] is uniquely
specified by the proper definition of a suitable family of phase-space LP's.
Eq.(\ref{Eq.2}) also provides the connection between Lagrangian and Eulerian
viewpoints. In fact the Eulerian pdf, $f(\mathbf{x},t),$ is simply obtained
from Eq.(\ref{Eq.2}) by letting $\mathbf{x}=\mathbf{x}(t)$ in the same
equation. As a result, the Eulerian and Lagrangian formulations of IKT, and
hence of the underlying moment (i.e., fluid) equations, are manifestly
equivalent.

\section{The exact dynamics of ideal tracer particles}

Let us now analyze in detail the equations of motion for ideal tracer
particles immersed in an incompressible thermofluid described by INSFE [Eqs.(%
\ref{1})-(\ref{4})] . First, it must be remarked that two types of forces
can in principle be present: a) a volume force, acted by the fluid (the same
one which is responsible of the phase-space Lagrangian dynamics); b)
particle-localized forces, such as the gravitational pull, acting directly
on the tracer particle. In particular, regarding the first one, its specific
form depends on the assumed pdf to be associated to the fluid. As discussed
elsewhere \cite{Tessarotto2009} its choice depends closely on the type of
fluid to be considered, i.e., deterministic or stochastic (as appropriate to
describe turbulent flows). In particular, the position $f\equiv f_{M}(%
\mathbf{x},t),$ with $f_{M}(\mathbf{x},t)$ defined by Eq.(\ref{Maxwellian})
is suitable for the description of deterministic, i.e., non-turbulent flows.
In such a case the appropriate form of the equation is obtained from Eqs.(%
\ref{F-1A})-(\ref{A}). Instead, the general case, in which one allows $f\neq
f_{M}(\mathbf{x},t),$ is provided by Eqs. (\ref{F-1A}) with (\ref{F-2}) and (%
\ref{F-3}).

Let us now assume that $m_{P}\neq m_{F},$ $m_{F}$ denoting the mass of the
displaced fluid. In view of the IKT approach the equations of motion depend
necessarily on the form of the pdf $f(\mathbf{x},t),$ and hence describe in
this sense \emph{the conditional phase-space dynamics}$.$ To construct the
equation of motion for an ideal tracer particle of arbitrary mass, let us
now assume, for definiteness, that the sole particle-localized force acting
directly on the tracer particle is produced by the gravitational pull. \emph{%
The equation of motion for an ideal tracer particle of mass} $m_{P}$ reads
simply in such a case:%
\begin{equation}
m_{P}\left[ \frac{d}{dt}\mathbf{v}(t)-\mathbf{g}\right] =m_{F}\left[ \mathbf{%
F}(\mathbf{x,}t;f)-\mathbf{g}\right] .
\label{equationtracer-dynamics-general}
\end{equation}%
which describes the conditional dynamics of an ideal tracer particle
immersed in an incompressible thermofluid described by a pdf $f(\mathbf{x}%
,t).$ We stress that the form of the pdf depends on the specific assumptions
made on the fluid \cite{Tessarotto2009}. The interpretation of this equation
is as follows. The terms on the l.h.s. represent the "inertial" and
gravitational forces acting on the tracer particle. Instead, all the terms
on the r.h.s. represent the volume force acting responsible for the
phase-space Lagrangian motion. The physical interpretation of the various
contributions appearing in the volume force [see Eqs.\emph{(}\ref{F-1A}),(%
\ref{F-2}) and (\ref{F-3})] is made transparent by representing them in
terms of the vector fields $m_{F}\mathbf{F}_{0}(\mathbf{x,}t;f)$ and $m_{F}%
\mathbf{F}_{1}(\mathbf{x,}t;f),$ to be interpreted as \emph{mean-field
forces.} One obtains in fact, in particular:%
\begin{equation}
m_{F}\left[ \mathbf{F}_{0}(\mathbf{x,}t;f)-\mathbf{g-}\frac{v_{th}^{2}}{%
2p_{1}}\mathbf{\nabla }p_{1}\right] \equiv m_{F}\mathbf{F}_{H}\mathbf{+}m_{F}%
\mathbf{D}(\mathbf{\mathbf{x,}}t),
\end{equation}%
It is immediate to prove that the terms in the first equation take into
account the \emph{fluid }and \emph{convective forces} $m_{F}\mathbf{F}_{H}$
and $m_{F}\mathbf{D}(\mathbf{\mathbf{x,}}t),$ with $\mathbf{F}_{H}$ and $%
\mathbf{D}(\mathbf{\mathbf{x,}}t)$ denoting respectively the fluid
acceleration (\ref{fluid acceleration}) and the convective term defined
above [see Eq.(\ref{D})]. In a similar way one can show that, in case $%
m_{P}\neq m_{F},$ Eq.(\ref{equationtracer-dynamics-general}) recovers also
the buoyancy force denbsity $\left( m_{P}-m_{F}\right) \mathbf{g}$ pointed
out by Maxey and Riley \cite{Maxey}. Finally, for the sake of comparison,
let us consider the case of an isothermal fluid and require - consistent
with Eq.(\ref{assumption}) - that locally in the extended phase-space $%
\Gamma \times I$ condition (\ref{assumption}) \ holds at time $t=t_{o}$. The
validity of previous tracer-particle equations [i.e.,\ in particular the MR
(Maxey-Riley) equation (\ref{MR equation})] \emph{requires that the
inequality (\ref{assumption}) holds} \emph{for all times} ($t\in I$)$.$ In
particular, one can prove that in such a limit our Eq.(\ref%
{equationtracer-dynamics-general}) agrees with the MR equation (\ref{MR
equation}). In fact, it yields%
\begin{equation}
m_{P}\frac{d}{dt}\mathbf{v}(t)\cong m_{F}\left. \frac{D\mathbf{V}(\mathbf{x}%
,t)}{Dt}\right\vert _{\mathbf{x=r}(t)}+m_{F}\mathbf{F}_{1}-\left(
m_{P}-m_{F}\right) \mathbf{g},  \label{APPROX-1}
\end{equation}%
where, in validity of (\ref{assumption}), it follows that the pressure
mean-field force can be approximated as $m_{F}\mathbf{F}_{1}\cong -\frac{1}{2%
}m_{F}\left. \frac{d}{dt}\left\{ \mathbf{v}(t)-\mathbf{V}(\mathbf{x}%
(t),t)\right\} \right\vert _{\mathbf{x=r}(t)}.$ On the other hand,\emph{\ a
serious objection }\ to all of the previous tracer-dynamics equations\emph{\
is provided by the possible violation of the asymptotic condition (\ref%
{assumption}), which may not be uniformly fulfilled in the set} $I.$ This
occurs, manifestly, if an initial condition of the type $\left\vert \mathbf{%
\mathbf{u}}(t_{o})\right\vert \sim \left\vert \mathbf{V}(\mathbf{r}(t_{o})%
\mathbf{,}t_{o})\right\vert $ is imposed on a tracer particle (this
requirement is not physically unreasonable since, in principle, tracer
particles might be injected in a fluid with arbitrary initial velocities).\
However, even if initially (at $t=t_{o}$) one requires the validity of (\ref%
{assumption}) in general it may well be also that $\left\vert \mathbf{%
\mathbf{u}}(t)\right\vert \sim \left\vert \mathbf{V}(\mathbf{r}(t)\mathbf{,}%
t)\right\vert $ at some later time ($t>t_{o}$). This can be achieved, for
example, even imposing the initial condition $\left\vert \mathbf{\mathbf{u(}}%
t_{o})\right\vert =0.$ The result is a consequence of Eq.(\ref%
{equationtracer-dynamics-general}). Indeed one can prove that,
even imposing the initial condition $\left\vert
\mathbf{u}(t_{o})\right\vert =0,$ generally $\left\vert
\mathbf{u}(t)\right\vert \neq 0,$ with $\left\vert
\mathbf{\mathbf{u}}(t)\right\vert $ not satisfying (\ref
{assumption}). In other words, ideal tracer particles having
initially the same local velocity of the fluid may develop in time
a finite relative velocity. This implies
that, generally, the full exact tracer-dynamics equation, i.e., Eq.(\ref%
{equationtracer-dynamics-general}), should be used, instead of the
asymptotic approximation indicated above [see Eq.(\ref{APPROX-1})].

\section{Concluding remarks}

In this paper the phase-space Lagrangian dynamics has been determined as
appropriate for an incompressible thermofluid described by a suitable set of
fluid equations [INSFE, see Eqs.(\ref{1})-(\ref{4})]. We have shown that,
based on the formulation of a restricted phase-space inverse kinetic theory,
the phase-space Lagrangian dynamics can be uniquely established. The
governing equations which determine the phase-space Lagrangian trajectories
(LP's) are found to depend functionally on the pdf [$f(\mathbf{x},t)$], to
be uniquely associated to the fluid by means of the IKT here adopted. \ In
particular, the theory permits to advance uniquely in time $f(\mathbf{x},t)$
and in terms of the same pdf also the complete set of fluid fields which
describe the fluid. This feature is of fundamental importance in turbulence
theory (see \cite{Tessarotto2009}).

As a further consequence, the dynamics of ideal tracer particles (i.e., for
which the perturbations of the fluid fields produced by the same\ particles
are negligible) is established. Remarkably, this result overcomes
limitations of customary ideal tracer-dynamics equations [see in particular
the Maxey and Riley equation \cite{Maxey} given by Eq.(\ref{MR equation})].%
\emph{\ }All of these equations are, actually, in disagreement
with the present theory for finite particle relative velocities.
The basic new result is represented by
Eq.(\ref{equationtracer-dynamics-general}) which describes,
\emph{for arbitrary initial velocity}, the conditional dynamics of
an ideal tracer particle in an incompressible thermofluid
described by a suitable pdf $f$.

\textbf{ACKNOWLEDGEMENTS} 

Work developed in cooperation with the CMFD Team, Consortium for
Magneto-fluid-dynamics (Trieste University, Trieste, Italy).\ Research
partially performed in the framework of the COST Action P17 (EPM, \textit{%
Electromagnetic Processing of Materials}), the GDRE (Groupe de Recherche
Europ\'{e}en) GAMAS and the MIUR (Italian Ministry of University and
Research) PRIN Programme: \textit{Modelli della teoria cinetica matematica
nello studio dei sistemi complessi nelle scienze applicate}.



\end{document}